\documentclass[aps,prr,floatfix,twocolumn,longbibliography,notitlepage,superscriptaddress,10pt]{revtex4-2}
\usepackage{xcolor}
\usepackage{grffile}
\usepackage{amsmath, amsthm, amssymb}
\usepackage[normalem]{ulem}
\usepackage{bm}
\usepackage{setspace}
\usepackage{grffile}
\usepackage{graphicx}   
\usepackage{verbatim}   
\usepackage{color}      
\usepackage{orcidlink}
\usepackage{hyperref}   
\usepackage[normalem]{ulem}
\usepackage{enumitem} 
\usepackage{dsfont} 
\usepackage{hyperref}
\hypersetup{colorlinks,linkcolor=blue,urlcolor=blue,citecolor=blue}

\newcommand{\bra}[1]{\langle #1|}
\newcommand{\ket}[1]{|#1\rangle}
\newcommand{\average}[1]{\langle #1\rangle}

\begin{document}

\title{Nonstabilizerness generation in a multiparticle quantum walk}
\author{C\u at\u alin Pa\c scu Moca\,\orcidlink{0000-0002-1299-399X}}
\email{mocap@uoradea.ro}
\affiliation{Department of Physics, University of Oradea,  RO-410087 Oradea, Romania}
\affiliation{MTA-BME Lend\"ulet ``Momentum'' Open Quantum Systems Research Group, Institute of Physics, Budapest University of Technology and Economics,
M\H uegyetem rkp. 3., H-1111, Budapest, Hungary}
\author{Doru Sticlet\,\orcidlink{0000-0003-0646-1978}}
\email{doru.sticlet@itim-cj.ro}
\affiliation{National Institute for R\&D of Isotopic and Molecular Technologies, 67-103 Donat, RO-400293 Cluj-Napoca, Romania}
\author{Bal\'azs D\'ora\,\orcidlink{0000-0002-5984-3761}}
\affiliation{Department of Theoretical Physics, Institute of Physics, Budapest University of Technology and Economics, M\H{u}egyetem rkp.~3, H-1111 Budapest, Hungary}
\affiliation{MTA-BME Lend\"ulet ``Momentum'' Open Quantum Systems Research Group, Institute of Physics, Budapest University of Technology and Economics,
M\H uegyetem rkp. 3., H-1111, Budapest, Hungary}
\author{Angelo Valli\,\orcidlink{0000-0002-0725-2425}}
\affiliation{Department of Theoretical Physics, Institute of Physics, Budapest University of Technology and Economics, M\H{u}egyetem rkp.~3, H-1111 Budapest, Hungary}
\affiliation{HUN-REN-BME-BCE Quantum Technology Research Group, Mûegyetem rkp. 3., H-1111 Budapest, Hungary}
\author{Dominik Szombathy\,\orcidlink{0009-0006-5224-245X}}
\affiliation{Department of Theoretical Physics, Institute of Physics, Budapest University of Technology and Economics, M\H{u}egyetem rkp.~3, H-1111 Budapest, Hungary}
\affiliation{Nokia Bell Laboratories, Nokia Solutions and Networks Kft, B\'okay J\'anos u. 36-42, H-1083 Budapest, Hungary}
\author{Gergely Zar\'and\,\orcidlink{0000-0001-7522-1826}}
\affiliation{Department of Theoretical Physics, Institute of Physics, Budapest University of Technology and Economics, M\H{u}egyetem rkp.~3, H-1111 Budapest, Hungary}
\affiliation{HUN-REN—BME Quantum Dynamics and Correlations Research Group,
 Budapest University of Technology and Economics, M\H{u}egyetem rkp.~3., H-1111 Budapest, Hungary}
\date{\today}

\begin{abstract}
	We investigate the generation of nonstabilizerness, or \emph{magic}, in a multiparticle quantum walk by analyzing the time evolution of the stabilizer R\'enyi entropy $M_2$.
	Our study considers both single- and two-particle quantum walks in the framework of the XXZ Heisenberg model with varying interaction strengths. 
	We demonstrate that the spread of magic follows the light-cone structure dictated by the system's dynamics, with distinct behaviors emerging in the easy-plane ($\Delta < 1$) and easy-axis ($\Delta > 1$) regimes. For $\Delta < 1$, magic generation is primarily governed by single-particle dynamics, while for $\Delta > 1$, doublon propagation dominates, resulting in a significantly slower growth of $M_2$. 
	Furthermore, the magic exhibits logarithmic growth in time for both one and two-particle dynamics.
	Additionally, by examining the Pauli spectrum, we show that the statistical distribution of level spacings exhibits Poissonian behavior, independent of interaction strength or particle number. 
	Our results shed light on the role of interactions on magic generation in a many-body system.
\end{abstract}
\maketitle


\section{Introduction} 
The notion of nonstabilizerness, or \emph{magic}~\cite{Leone2022}, has emerged as a key concept in quantum information theory, quantifying how far a quantum state deviates from the stabilizer states' subspace~\cite{Gottesman1998,Aaronson2004}. 
By measuring the distance from efficiently classically simulable stabilizer states, magic serves both as a fundamental resource for achieving quantum advantage~\cite{bao2022magic,Oliviero2022,zhang2024quantum,szombathy2024,gidney2024magic,Niroula2024}, and as a tool to understand the complexity of quantum correlations in many-body systems~\cite{Rattacaso2023,tarabunga2024critical,gu2024}.

Quantum walks~\cite{kadian2021quantum,Childs.2009,venegas2012quantum} provide an ideal platform to study the generation of magic in a controlled setting. 
In a single-particle quantum walk, for instance, the dynamics of a single spin-flip in a one-dimensional spin chain can be mapped onto a single-particle quantum walk, with the initial state chosen as a stabilizer state~\cite{Aaronson2004}. 
As the system evolves under unitary dynamics, interference effects cause the Pauli spectrum~\cite{Beverland2020, Turkeshi2025a} to delocalize, leading to the growth of stabilizer R\'enyi entropy. 
This process reflects the wavefunction's exploration of Hilbert space regions beyond the stabilizer subspace.
From this perspective, quantum walks provide a minimal setting in which the emergence of non-Clifford complexity can be directly linked to spatial and interference-driven delocalization. 
Unlike generic many-body dynamics, quantum walks allow a clear mapping between real-space propagation and the growth of magic while. 
Thus they provide a quantitative probe into how coherent transport and interference drive a state away from efficiently classically simulable dynamics.

In this work, we investigate the dynamics of magic~\cite{Rattacaso2023,zhang2024quantum,Tirrito2024,Bejan2024,Turkeshi2025} in single- and multiparticle quantum walks~\cite{childs2013universal,Cai2021,Ostahie2023,Mittal2025}.
We employ the Pauli operator basis to analyze the evolution of stabilizer states under unitary dynamics, and explore the influence of interactions on magic generation~\cite{szombathy2024,szombathy2025}.

\paragraph*{Stabilizer R\'enyi entropy and the Pauli spectrum.} The concept of magic, or nonstabilizerness, can be rigorously defined using the stabilizer R\'enyi entropy~\cite{Leone2022}, which quantifies the deviation of a quantum state from the stabilizer subspace. 
For a pure quantum state $\ket{\psi}$, the stabilizer R\'enyi entropy $M_2$ is defined as  
\begin{equation}  
M_2 = -\log_2 \sum_P c_P^4,  
\end{equation}  
where the coefficients $c_P$ are determined from the expansion of the density matrix $\rho = \ket{\psi}\bra{\psi}$ in the Pauli operator basis,  
\begin{equation}  
\ket{\psi}\bra{\psi} = \frac{1}{2^N} \sum_P c_P P,  
\end{equation}  
with $P$ denoting elements of the $N$-qubit Pauli group~\cite{Poulin2005,Garcia2023,arab2024lecture} and $c_P = \mathrm{Tr}(P \ket{\psi}\bra{\psi}) = \bra{\psi} P\ket{\psi}$. 
In the context of the  quantum walk that we address, the unitary time evolution of the initial state  
generates interference effects in the Pauli spectrum as the wavefunction propagates through the chain. 
This results in the delocalization of the Pauli coefficients $c_P$, leading to the growth of $M_2$. 
As the  N-qubit Pauli group consists of $4^L$ elements, there are  $4^L$  coefficients $c_P$, that form the \emph{Pauli spectrum}.
In practice, computing the full spectrum becomes 
computationally expensive for large $L$ due to the exponential growth of the Pauli group size. 
Efficient numerical methods for evaluating the stabilizer entropy use  
Monte Carlo sampling techniques~\cite{Tarabunga2023,Liu2025}, or  for particular states such as generalized Rokhsar-Kivelson, the magic $M_2$ can be evaluated analytically~\cite{tarabunga2024magic}. 
Recent advancements also explore tensor network methods~\cite{Haug2023,Lami2023,Tarabunga2024,Chen2024,Mello2024,Frau2024,Lami2025}, which allow for the efficient calculation of $M_2$ in systems with moderate entanglement, where typical bond dimensions are small enough. 
In the quantum walk problems discussed here, we evaluate $M_2$ both analytically, using the method
proposed in Ref.~\cite{tarabunga2024magic}, and numerically through the matrix product states (MPS) formalism~\cite{Haug2023,Tarabunga2024} implemented using the ITensors.jl library~\cite{Fishman2022, Fishman2022a}. 
In the single-particle quantum walk, the bond dimension is always restricted to $M=1$, 
whereas for the many-body (two-particle) quantum walk, a good approximation of the wavefunction's time evolution 
is obtained for a bond dimension of $M <10$.

\section{Single-particle quantum walk} 
The one-dimensional XX chain~\cite{Lieb1961,Katsura1962} provides the simplest possible framework to study the dynamics of a single-particle quantum walk by considering the spin-1/2 degrees of freedom as qubits. 
The Hamiltonian of the XX chain, expressed in terms of qubits, is given by
\begin{equation}  
{\cal H}_0 = -\frac{J}{4}\sum_{j=-L/2}^{L/2-1} \left( X_j X_{j+1} + Y_j Y_{j+1} \right),  
\label{eq:H_XX}
\end{equation}  
where $J$ is the coupling strength (associated with the energy unit), $X_j$ and $Y_j$ are the Pauli-$X$ and Pauli-$Y$ operators acting on the $j$th qubit, and open  boundary conditions are typically assumed. The single-particle excitation corresponds to a single spin flip propagating through the chain, which can be mapped to a quantum walk on a lattice. The initial quench generates a light-cone propagation with a Lieb-Robinson velocity $v_F = J$~\cite{nachtergaele2010lieb,Chen_2023}. We work throughout in natural units $\hbar=1$ and with lattice constant $a=1$.

The initial state constructed in the computational local basis $\ket{j} = \{ \ket{0}, \ket{1}\}$, 
with all qubits being in the  $\ket{0}$, except for one qubit in the middle of the chain, associated with $j=0$, which is flipped to the state $\ket{1}$,
\begin{equation}
\ket{\psi(0)} = \ket{0 \, 0 \, \cdots \, 0 \, 1 \, 0 \, \cdots \, 0}.
\label{eq:psi0_SP}
\end{equation} 
This state represents a localized, single-particle excitation, which can propagate through the chain under the action of the Hamiltonian \eqref{eq:H_XX}.
The time-dependent wavefunction,  $ \ket{\psi(t)} = e^{-i{\cal H}_0t} \ket{\psi(0)}$, can be expanded in the local basis. 
The probability amplitude of finding the excitation at site $j$ at time $t$ is given by $\psi_j(t) = \average{j|\psi(t)}$. For the initial state considered here, the coefficients of the wavefunctions are expressed in terms of the Bessel functions~\cite{Antal1999, Eisler2013},  $\psi_k(t) = i^{-k} J_k(v_F t)$.

By construction, the initial state~\eqref{eq:psi0_SP} is a stabilizer state with zero magic,
as quantified using the stabilizer R\'enyi entropy, $M_2(\ket{\psi(0)}) =0$. 
As the single-particle excitation propagates through the chain, the wavefunction $\ket{\psi(t)} $ typically spreads and explores regions of Hilbert space beyond the stabilizer subspace, as the unitary evolution induced by the Hamiltonian~\eqref{eq:H_XX} drives the generation of nonstabilizerness (magic).
Using the approach outlined in Ref.~\cite{tarabunga2024magic}, magic is determined from the wave-function coefficients as
\begin{equation}
	2^{-M_2} = \sum_{ijkl} \psi_j \psi_k \psi_l \psi^*_m
	\psi_{[jkl]}
	\psi^*_{[jkm]}
	\psi^*_{[jlm]}
	\psi^*_{[klm]},
\end{equation}
with the time argument of $\psi_j$ omitted for brevity, and
\begin{equation}
	\psi_{[jkl]} = \psi_j\delta_{kl}+\psi_k\delta_{jl}+\psi_l\delta_{jk}-2\psi_j\delta_{jk}\delta_{jl}.
\end{equation}
By performing the summations over the site indices, one finds the stabilizer R\'enyi entropy
\begin{equation}\label{eq:magic_coeffs}
	2^{-M_2} = \sum_j \big[-6|\psi_j|^8 + 
	\sum_k \big( 6|\psi_j|^4|\psi_k|^4 
	+
	\psi_j^4 \psi_k^{*4}
	\big)
	\big].
\end{equation}
Thus, substituting the explicit Bessel function representation of single-particle wavefunction coefficients, yields an exact expression for $M_2$ as
\begin{equation}
	M_2 = -\log_2\left[ \sum_{j,k} (7-6\delta_{jk})J_j^4(v_F t)J_k^4(v_F t)\right],
	\label{eq:magic_SP}
\end{equation}
with $j,k$ running over the lattice sites. While this expression cannot be evaluated analytically, it has an accurate asymptotic form in the limits of large $t$ and $L$~\cite{Stoyanov1987, Martin2007},  
\begin{equation}  
	M_2 \approx -\log_2\left[\frac{7}{\pi^4 (v_F t)^2} (\ln (v_Ft) + 5 \ln (2) + \gamma)^2\right],  
	\label{eq:magic_SP_asympt}
\end{equation}  
where $\gamma$ is the Euler-Mascheroni constant. This result indicates that, in the thermodynamic limit, the magic exhibits asymptotic growth as $\sim \log_2(v_F t)$ at large $t$, with a subleading correction of the form $\sim \log_2[\ln (v_F t)]$.  In the limit $t \to 0$, the initial entropy increases quadratically, following $M_2 \propto (v_F t)^2$. 

\begin{figure}[t!]
	\begin{center}
	 \includegraphics[width=0.9\columnwidth]{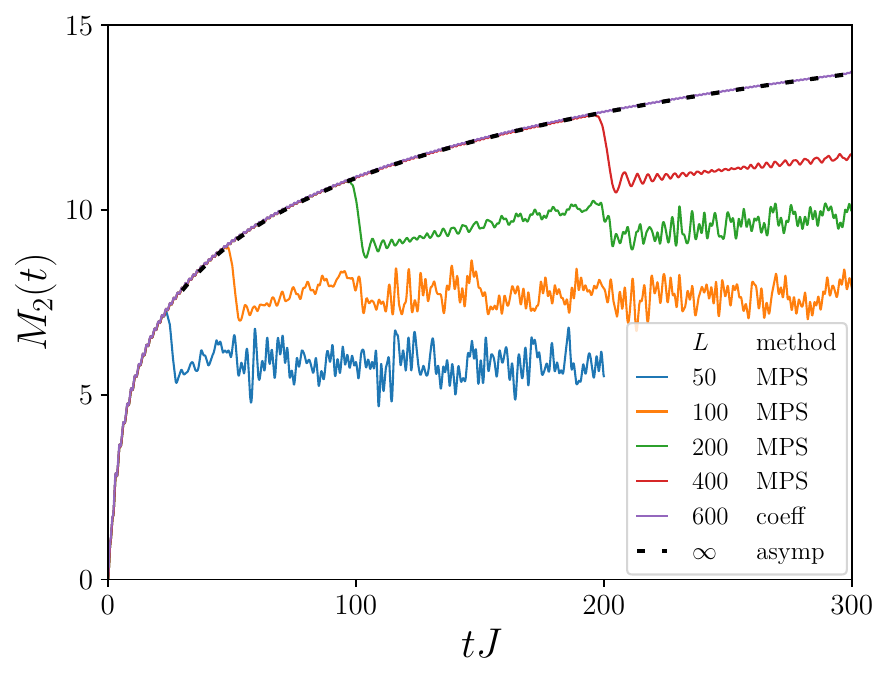}
	 \caption{Time evolution of the stabilizer R\'enyi entropy $M_2$ in the single-particle quantum walk, shown as a function of time for different system sizes. 
	 The MPS data for $L\in\{50,100,200,400\}$~\cite{Haug2023, Tarabunga2024} is compared against (coeff) results from wave-function coefficients~\eqref{eq:magic_SP} for $L=600$, and (asymp) the asymptotic expression~\eqref{eq:magic_SP_asympt}.
	 }
	 \label{fig:Magic_SP}
	\end{center}
\end{figure}

The typical behavior of $M_2$ is illustrated in Fig.~\ref{fig:Magic_SP}, where the result of Eq.~\eqref{eq:magic_SP} is compared to both the analytical formula~\eqref{eq:magic_SP_asympt}, and the numerical data obtained via the matrix product state approach~\cite{Haug2023,Tarabunga2024,Mello2024}. 
Expression \eqref{eq:magic_SP} remains accurate until boundary effects become relevant. 
Once edge reflections occur, $M_2$ undergoes a sharp drop to specific values, followed by time-dependent finite-size fluctuations.  

The generation of magic is constrained within the light cone, following the causality imposed by the underlying quantum dynamics. 
$M_2$ grows as the wavefunction spreads through the system, but only within regions connected by the maximal propagation velocity. 
Outside the light cone, correlations and non-classical features remain negligible, indicating that magic is entirely created and dynamically reshaped within the accessible causal domain.

\section{Multiparticle quantum walk.}

\begin{figure}[t!]
	\begin{center}
	\includegraphics[width=0.9\columnwidth]{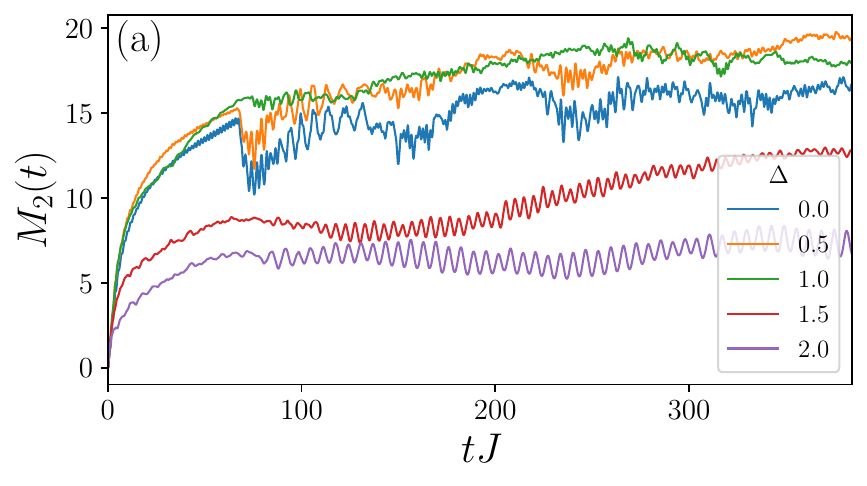}
	\includegraphics[width=\columnwidth]{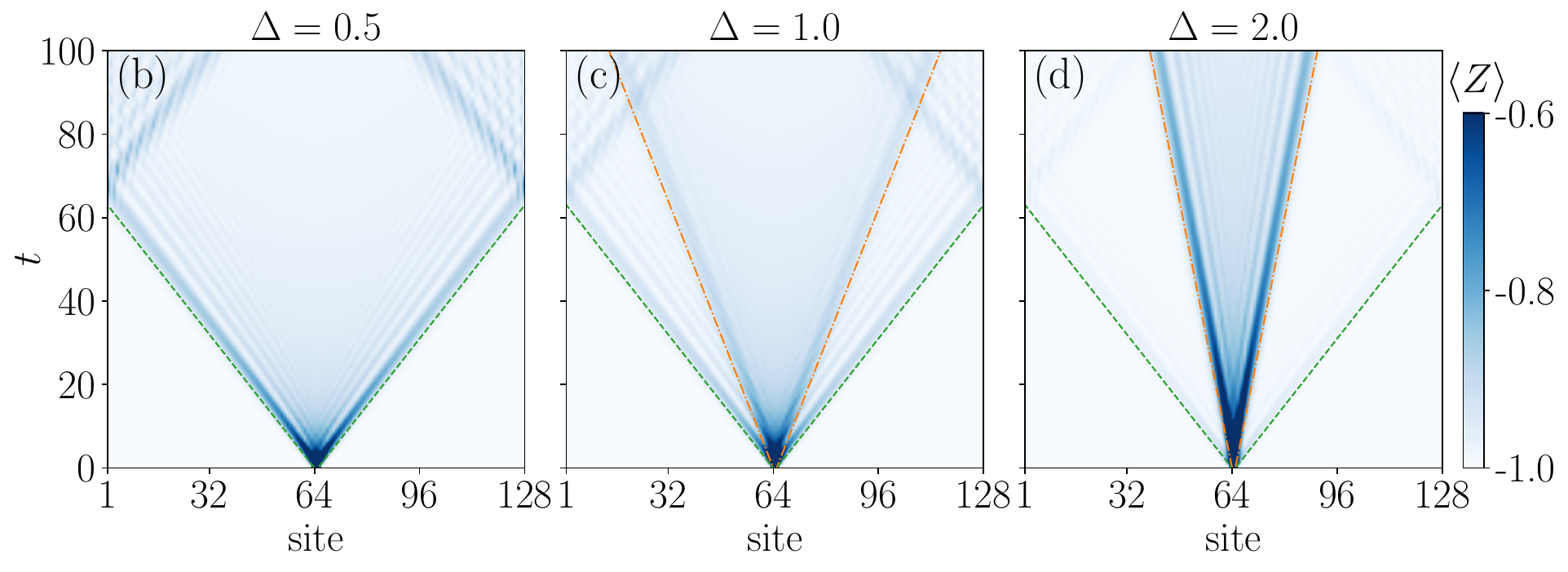}
	\caption{
	 (a) Time evolution of magic $M_2$ in the two-particle quantum walk for different interaction strengths $\Delta$, using the MPS approach. 
	 (b)--(d) Density plots showing the local magnetization dynamics $\langle Z\rangle$ for $\Delta = \{0.5, 1, 2\}$, obtained using exact diagonalization.
	 In each plot, the green lines indicate the boundaries of the single-particle light cone, while the orange lines mark the doublon light cone with $v^{(D)}_F=0.5J/\Delta$. 
	 The density intensity is truncated near $\langle Z\rangle=1$ for better visibility of the single-particle light cones at $\Delta\geq 1$. 
	 In all the panels the system size is fixed to $L=128$.
	}
	 \label{fig:Magic_2P}
	\end{center}
\end{figure}
\begin{figure}[t!]
\centering
\includegraphics[width=0.9\columnwidth]{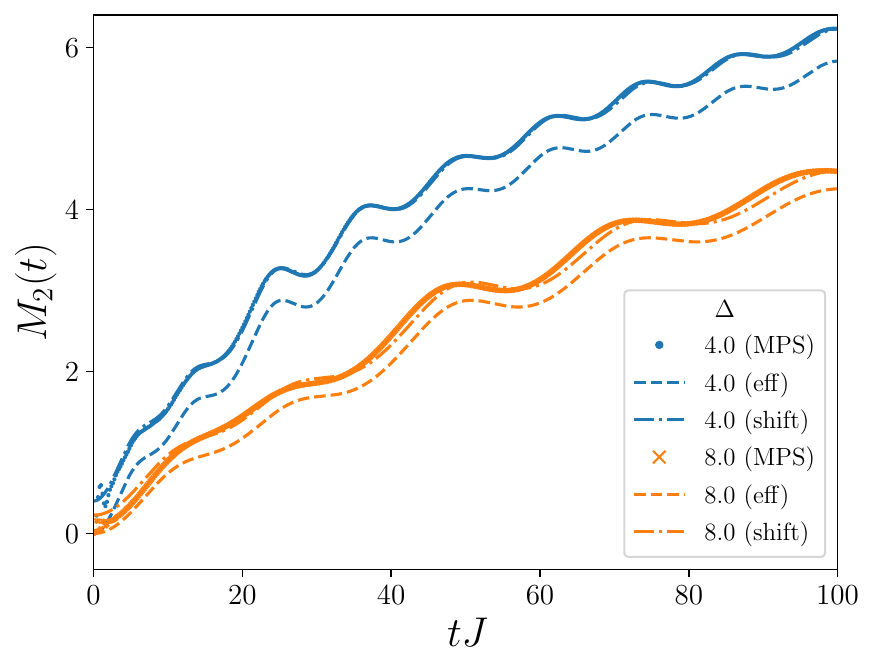}
\caption{The time evolution of $M_2$ for large $\Delta$ values. 
Solid lines show the total $M_2$ computed using the MPS algorithm, while dashed lines represent the doublon contribution. The dash-dotted lines represent the effective doublon contribution shifted by a constant, which reproduces at large $t$ the MPS results.
The system size is $L=128$ for MPS simulations. 
} 
\label{fig:Magic_doublon}
\end{figure}

In the following, we address the role of interactions by considering a scenario where two neighboring qubits are in the state $\ket{1}$ while the remaining qubits are in the state $\ket{0}$. 
To incorporate interactions, we modify the Hamiltonian~\eqref{eq:H_XX} by adding an interaction term and study the generation of magic under the unitary evolution governed by  the XXZ Heisenberg model as
\begin{equation}\label{eq:XXZ}
{\cal H} = {\cal H}_0 - \frac{\Delta J}{4} \sum_{j=-L/2}^{L/2-1} Z_j Z_{j+1},
\end{equation}
where $\Delta$ represents the anisotropy parameter, and the $Z_j Z_{j+1}$ term accounts for nearest-neighbor interactions along the chain. 

The time evolution of magic is shown in Fig.~\ref{fig:Magic_2P}(a) for different values 
of $\Delta$. The same figure also presents the time evolution of the magnetization 
$\average{Z}(x,t)$ along the chain, where the formation of a light cone is clearly visible for 
three distinct values of $\Delta$.  

For $\Delta < 1$, in the easy-plane regime, interactions play a small role on the dynamics, and the underlying single-particle behavior is dominant. 
This is evident in Fig.~\ref{fig:Magic_2P}(b), where a single light-cone velocity, $v_F=J$, emerges. 
In this regime, the magic follows a similar time dependence trend, growing asymptotically as $M_2(t) \sim \log_2 (v_F t)$. 
A characteristic drop in $M_2$ is observed when the light cone reflects off the system boundaries.

For $\Delta > 1$ in the easy-axis regime, single-particle dynamics are primarily governed by the motion of a 
composite excitation formed by the two adjacent $|1\rangle$ qubits. 
This arises from energy considerations: When $\Delta > 1$, the initial state with paired spins occupies a high-energy configuration, separated from other 
states by a gap proportional to $\Delta$. 
Energy conservation then ensures that the two spins move together as a bound composite object known as a \emph{doublon}.

This behavior becomes evident after applying a Jordan-Wigner transformation from qubits to spinless fermions, followed by a perturbative expansion in $J$~\cite{Ostahie2023}. 
The resulting doublon, composed of two fermions, behaves as a boson and its dynamics can be described by an effective Hamiltonian, 
\begin{equation}\label{eq:h_eff}
\mathcal H_{\rm eff}\simeq \sum_j
-\frac{J_{\rm eff}}{4}\left[X^{(D)}_jX^{(D)}_{j+1}+Y^{(D)}_jY^{(D)}_{j+1}\right]
+h_{\rm eff}Z_j^{(D)},
\end{equation}
where the effective hopping is given by $J_{\rm eff} = J/2\Delta$, and the effective magnetic field by $h_{\rm eff} = (\Delta J + J_{\rm eff})/2$. 
Here, the superscript $D$ added to the Pauli matrices denotes doublon operators.  

As a result, the Lieb-Robinson velocity for doublons is renormalized by interactions, given by $v^{(D)}_F(\Delta) = J_{\rm eff}$, making doublon dynamics the primary contributor to magic. 
This is clearly illustrated in Fig.~\ref{fig:Magic_doublon}, where the magic associated with doublon dynamics, computed using Eq.~\eqref{eq:magic_SP} with $v_F \mapsto v_F^{(D)}$, is compared to the total magic computed using the MPS approach. 
As expected, for large $\Delta \gg 1$, the contribution from single-particle dynamics is negligible, while doublon motion, governed by $H_{\rm eff}$, accounts for the majority of magic generation. At large $t$, the doublon magic, shifted by a constant that diminishes as $\Delta$ increases, closely reproduces the total magic observed in MPS simulations.

The formation of magic is strictly confined within the respective light cones, reflecting the causal structure imposed by the system's dynamics. For $\Delta < 1$, magic emerges predominantly through single-particle propagation, expanding at $v_F=J$. 
In contrast, for $\Delta > 1$, magic develops mostly within the slower-moving doublon light cone, where its growth is dictated by the effective hopping dynamics of the bound pairs. 
At the intermediate point $\Delta = 1$, both types of excitations contribute to the spread of magic, leading to a more intricate interplay between single-particle and doublon-mediated processes. 
\begin{figure}[t!]
	\begin{center}
		\includegraphics[width=0.9\columnwidth]{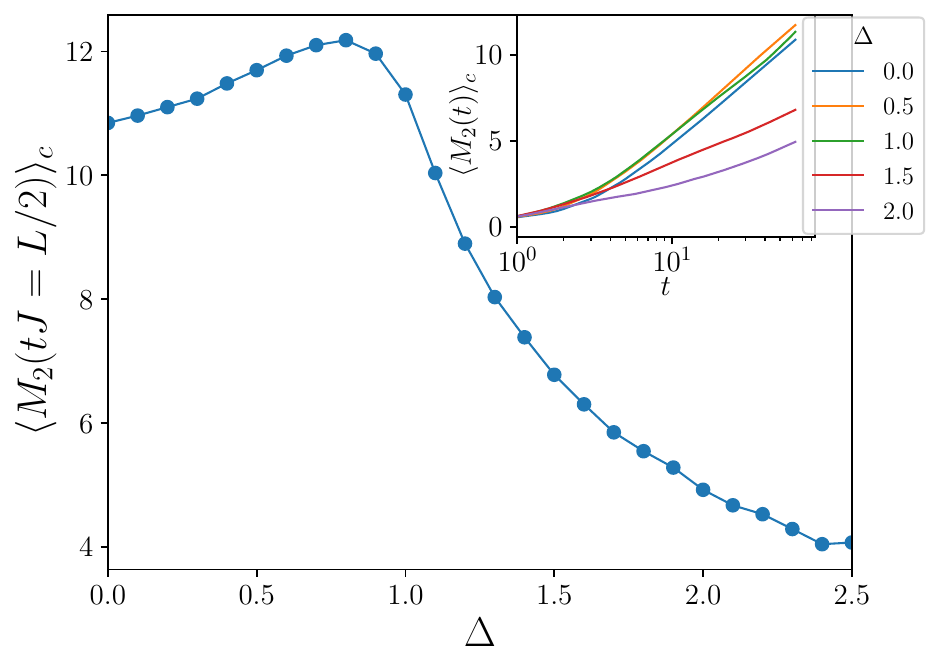}
		\caption{The cumulative average of magic as a function of $\Delta$ for the two-particle quantum walk, measured before the single-particle light cone reaches the system boundary at $tJ \approx L/2$. 
		The inset shows the logarithmic growth of $\langle M_2(t)\rangle_c$ at large $t$ for several $\Delta$ values.
		The system size is $L=128$.
		}
		\label{fig:Magic_average}
	\end{center}
\end{figure}

To analyze the magic in the early-time regime, before the single particle reaches the system boundaries at 
$t \simeq L/2$, we consider the cumulative average $\langle M_2(t) \rangle_c = \int_0^t dt' M_2(t') / t$, which shares the same time-scaling as $M_2(t)$, but smoothened fluctuations. 
The results, shown in Fig.~\ref{fig:Magic_average}, demonstrate the decrease in cumulative magic as dynamics is dominated by doublons forming and propagating at $\Delta>1$.
This phenomenon is associated with the increasing effective mass of the doublon, which slows its propagation and reduces its contribution to magic generation.
Moreover, for all $\Delta$, the (cumulative) magic obeys a logarithmic growth in time [Fig.~\ref{fig:Magic_average}(inset)], similar to the single-particle result in Eq.~\eqref{eq:magic_SP_asympt}, albeit with different prefactors and effective velocities.
This reinforces the previous observation that at $\Delta\gg1$ doublon magic generation is governed by Eq~\eqref{eq:magic_SP} with renormalized velocity, while at $\Delta<1$, single-particle physics dominates with $v_F=J$.
The logarithmic scaling in the intermediate regime is due to contribution of both single- and two-particle dynamics.
After multiple reflections from the system boundaries, the magic approaches a stationary regime in the 
long-time limit, characterized by significant finite-size fluctuations [see Fig.~\ref{fig:Magic_2P}(a)].
In this regime, the average magic density, $\langle M_2 \rangle / L$, while dependent on $\Delta$, remains relatively stable and with increasing $\Delta$, it gradually decreases. 

\section{Pauli spectrum}
\begin{figure}[t!]
	\begin{center}
	\includegraphics[width=\columnwidth]{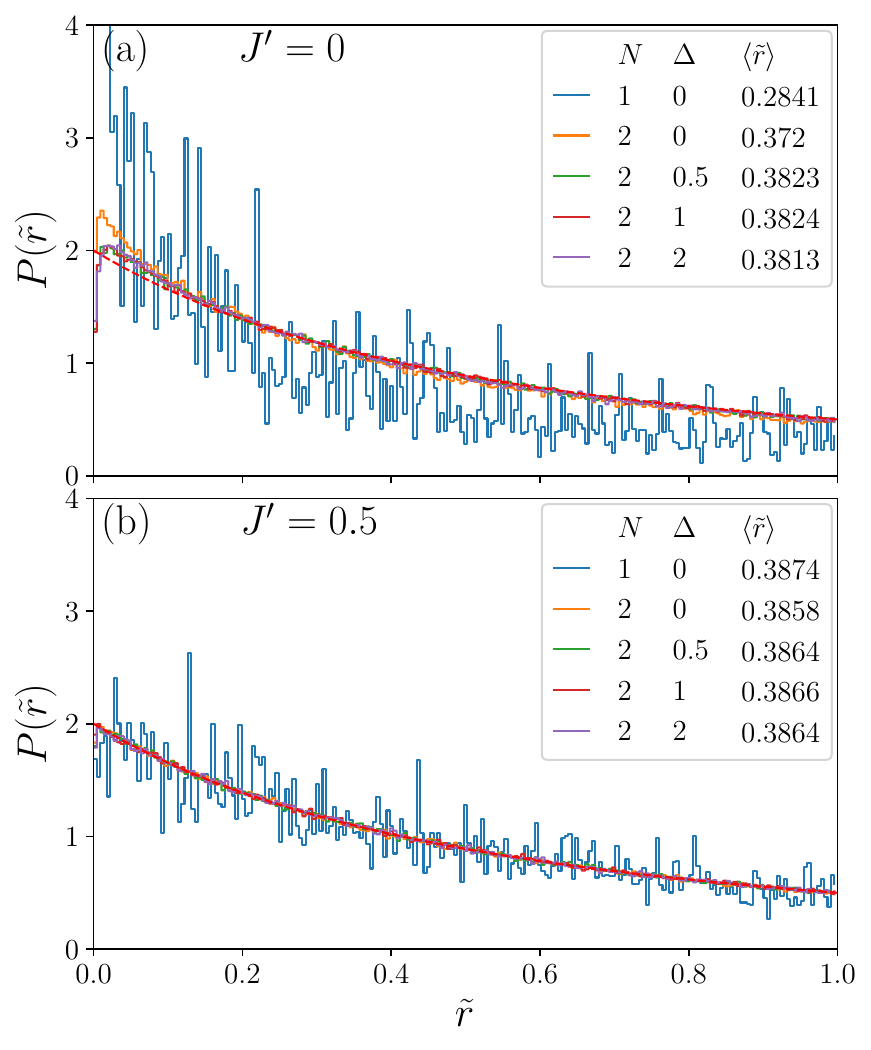}
	\caption{The distribution of the ratios $P(\tilde r)$ for the Pauli string values in the long-time limit for the single and two-particle quantum walk.
	The dynamics is governed by (a) the XXZ Hamiltonian~\eqref{eq:XXZ} or (b) the nonintegrable next-nearest-neighbor hopping XXZ model~\eqref{eq:H_nnn} with $J'=0.5J$.
	The red dashed line corresponds to the Poisson distribution, $P(\tilde r) =2/(1+\tilde r)^2\theta(\tilde r) \theta(1-\tilde r)$~\cite{Atas2013}. }
	 \label{fig:Pauli_spectrum_SP}
	\end{center}
\end{figure}
%
Next, we examine the Pauli spectrum in the context of the quantum walk. 
In both the single- and two-particle quantum walks, the initial state is a 
stabilizer state, meaning that at $t=0$, the Pauli spectrum is concentrated at three values, with coefficients $c_P$ taking values in $\{0, \pm 1\}$. 
Specifically, among the $4^L$ possible string values, $2^L(2^L - 1)$ are zero, while $2^{L-1}$ are equal to $1$, and $2^{L-1}$ are equal to $-1$.
As time evolves, the spectrum becomes increasingly dense, eventually reaching a long-time limit characterized by a mixture of a delta peak at zero and a continuous distribution, $ \rho(c_P) = A \delta(c_P) + \rho_c(c_P)$.
For instance, in a system of size $L=10$, out of the $4^{10}$ possible Pauli string values, only about 1\% remain nonzero in the long-time limit.  

To analyze the distribution of these nonzero values, we construct an ordered set of string values $c_n = \text{sort}(c_P)$ and examine the nearest-neighbor spacings $s_n = c_{n+1} - c_n$. 
From this, we compute the ratio  
\begin{equation}
\tilde{r}_n = \frac{\min(s_n, s_{n-1})}{\max(s_n, s_{n-1})} = \min\left(r_n, \frac{1}{r_n}\right),
\end{equation}  
where $r_n = s_n / s_{n-1}$. 
The probability distribution $P(\tilde{r})$ is then compared to traditional level-spacing distributions~\cite{wigner1967random,Atas2013}. 
By construction, $P(\tilde{r})$ is supported on the interval $[0,1]$.  

The results, shown in Fig.~\ref{fig:Pauli_spectrum_SP}, indicate that the distribution of ordered Pauli string ratios follows a Poissonian form, with an average value $\langle \tilde{r} \rangle$ that closely matches the theoretical prediction $2\ln 2 - 1 \approx 0.38629$~\cite{Atas2013}. 
As depicted in Fig.~\ref{fig:Pauli_spectrum_SP}(a), in the long-time limit, the level-spacing distribution remains unchanged regardless of whether the system undergoes a single- or two-particle quantum walk. Furthermore, it is insensitive to the interaction strength $\Delta$, suggesting a universal statistical behavior in the asymptotic regime.
In order to check the robustness of this finding, we have considered an extended XXZ model with next-nearest-neighbor interactions, thus breaking the model's integrability,
\begin{equation}\label{eq:H_nnn}
\mathcal H' = \mathcal H - \frac{J'}{4}\sum_{j=-L/2}^{L/2-2}
(X_iX_{i+2}+Y_iY_{i+2}).
\end{equation}
The magnetization dynamics shows similar ballistic behavior, generically with double-lightcone structure at any $\Delta$, due to the presence of distinct hoppings $J$ and $J'$.
The Pauli spectrum analysis in the long-time limit of stationary magic for short chains [from $L=4$ to $L=10$] reveals again a Poissonian distribution, with average spacing ratio close to the characteristic theoretical value $0.38629$ [Fig.~\ref{fig:Pauli_spectrum_SP}(b) for $L=10$ and $J'=0.5J$].
These findings suggest that Pauli eigenvalues evolve uncorrelated in the regime of stationary magic.


\section{Conclusions}

In this study, we investigated the dynamics of nonstabilizerness, or \emph{magic}, in single and multiparticle quantum walks by analyzing the time evolution of the stabilizer R\'enyi entropy $M_2$.
Our findings reveal that the generation and propagation of magic are intricately linked to the light-cone structure dictated by the system's dynamics. 

For single-particle quantum walks, we obtain an analytic expression for the magic spreading before the walker hits the boundary of the system. 
The stabilizer R\'enyi entropy exhibits slow, logarithmic growth in time.
For the multiparticle case, we consider as initial state a ferromagnetic background with two neighboring spins flipped in the opposite direction. 
Then, the system evolves according to the XXZ Heisenberg model.
In the easy-plane regime ($\Delta < 1$), magic spreads logarithmically with $v_F$ close to the single-particle Fermi velocity $v_F=J$, similarly to the single particle quantum walk case. 
This regime is indeed dominated by single-particle dynamics, where the growth of magic is primarily driven by the propagation of individual excitations. 
In contrast, in the easy-axis regime ($\Delta > 1$), the dynamics is governed by the motion of doublons---bound pairs of spins---resulting in a significantly slower growth of magic due to the increased effective mass of these composite excitations. 
After reflections at the system boundary, the magic tends to saturate, and exhibits finite-size fluctuations.

Furthermore, we examined the statistical properties of the Pauli spectrum in the magic-saturation regime and found that the level-spacing distribution follows a Poissonian form, with an 
average spacing ratio consistent with theoretical predictions. 
This universality holds regardless of the interaction strength $\Delta$ or the number of particles involved in the quantum walk, 
indicating a robust statistical behavior in the asymptotic regime.
Moreover, breaking the integrability of the model by introducing next-nearest-neighbor hopping did not affect the Poissonian nature of the Pauli spectrum.
Our results underscore the relationship between interactions and nonstabilizerness in many-body quantum systems, revealing the distinct roles of single-particle and doublon propagation in 
the generation of magic. 
These insights pave the way for future research into the effects of disorder~\cite{tarabunga2024magic}, external driving~\cite{Tirrito2024}, and other perturbations on the evolution of magic in quantum systems~\cite{Turkeshi2024,Sticlet2025,Tirrito2025a}. 

\begin{acknowledgments}
This work received financial support from CNCS/CCCDI-UEFISCDI, under projects No. 
PN-IV-P1-PCE-2023-0159, PN-IV-P1-PCE-2023-0987, PN-IV-P8-8.3-PM-RO-FR-2024-0059, and by the 
``Nucleu'' Program within the PNCDI 2022-2027, Romania, carried out with the support of MEC, 
project no.~27N/03.01.2023, component project code PN 23 24 01 04.
This research was also supported by the National Research, Development and Innovation Office - 
NKFIH within the Quantum Technology National Excellence Program (Project
No.~2017-1.2.1-NKP-2017-00001 and grant No. SNN139581), K134437, K142179 by the BME-Nanotechnology FIKP grant (BME FIKP-NAT), and the QuantERA `QuSiED' grant No.~10101773.
This work was also supported by the HUN-REN Hungarian Research Network through the Supported Research Groups Programme, HUN-REN-BME-BCE Quantum Technology Research Group (TKCS-2024/34).

\end{acknowledgments}




\bibliography{references}
\end{document}